\def\NIMA{{\em Nucl. Instrum. Methods} A}
\documentstyle[epsfig]{aipproc}
\begin{document}
%%%%%%%%%%%%%%%
\vspace*{-1.7cm}
\begin{flushright}
RU 02/E-10
\end{flushright}
%%%%%%%%%%%%%%%
\vglue .2in

\title{The CDF MiniPlug Calorimeters
\footnote{Contributed paper
to ``DPF 2002, Section: New Detector Technologies'', Williamsburg,
VA, May~24-28~2002}}
\author{Stefano Lami
\footnote{{\em E-mail address}: lami@physics.rockefeller.edu}}
\address{(CDF Collaboration) \\ \vspace{.2in}
The Rockefeller University, New York, NY 10021-6399, USA}
\maketitle
\pagenumbering{arabic}
\begin{abstract}
Two MiniPlug calorimeters,
designed to measure the energy and lateral position of particles 
in the (forward) pseudorapidity region of $3.6<|\eta|<5.2$ of the
CDF detector, have been recently installed as part of the Run II CDF upgrade
at the Tevatron $\bar pp$ collider.
They consist of lead/liquid scintillator read out by wavelength shifting
fibers arranged in a pixel-type towerless geometry suitable for
`calorimetric tracking'.
The design concept, the prototype performance and 
the final design of the MiniPlugs are here described.
A recent cosmic ray test resulted in a light yield of approximately 
100 pe/MIP, which exceeds our design requirements.
\end{abstract}
%
%%%%%%%%%%%%%%%%%%%%%%%%%%%%%%%%%%%%%%%%%%%%%%%%%%%%%%%%%%%%%%%%%%%%%%%%%%%%%%%
\section{Introduction}

A program of hard diffraction and very forward physics
proposed for CDF in Run II~\cite{PAC-proposal} 
requires the employment of two forward
`MiniPlug' calorimeters in the pseudorapidity region $3.6<|\eta|<5.2$
designed to measure the energy and lateral position of both electromagnetic and
hadronic showers. The MiniPlugs extend the pseudorapidity 
region covered by the Plug Upgrade calorimeters ($1.1<|\eta|<3.5$)
to the beam pipe.
Using the MiniPlug and the Plug Upgrade calorimeters 
to actually measure the width of the rapidity gap(s) produced 
in diffractive processes will allow extending the  Run I 
studies of the diffractive structure function to much lower values of the 
fraction $\xi$ of the momentum of the proton carried by the Pomeron.

The MiniPlugs are based on lead/liquid-scintillator
layers read out by wavelength shifting (WLS) fibers perpendicular to the 
lead plates and parallel to the 
proton/antiproton beams, arranged in a novel `towerless'
geometry (no boundaries between the fibers).
The centroid of the fiber pulse height provides the position 
of the shower initiating particle.
The desired number of fibers is then grouped and
viewed by a single pixel of a multi-anode photomultiplier 
tube (MAPMT), providing what in the following we will call for
simplicity a `tower'.

As interacting hadrons release on average one third of their energy
in the form of $\pi^{0}$'${\rm s}$,
 a short (few interaction lengths, $\lambda$)
calorimeter can be used  to measure the energy and position of both 
electrons/photons and hadrons.
A set of fibers which do not sample the first 24 radiation lengths
($X_0$)
of the detector can be used to tag hadrons.
 
\noindent\begin{minipage}[b]{2.5in}{
\vglue 0.2cm
The MiniPlugs have been installed downstream of the CDF Plug
Upgrade calorimeters within the central hole of the muon toroids
at a distance of 5.8 meters from the 
center of the detector (see fig. 1).
Each MiniPlug is housed in a cylindrical steel vessel 26$''$ 
(66 cm) in diameter and has a 5$''$ (12.7 cm) hole concentric with  
the cylinder axis to accommodate the beam pipe.
Due to the space constraints in the $z$-direction when 
the Plug is withdrawn to the ``open" position to service the central 
detectors,
the MiniPlug length is confined to 24$''$ (61 cm). 
Within this length the actual calorimeters are 32 $X_0$ and 
1.3 $\lambda$ deep, and have no hadron tagging fibers.}

The MiniPlug design combines a low cost construction with an efficient
and high-resolution position determination.
The design concept and MonteCarlo expectations are described
in section II.
\end{minipage}
\hspace*{.1in}
\begin{minipage}[t]{3.1in}{ 
\vspace*{-12.cm}
\centering
\vglue -.1in
\epsfig{file=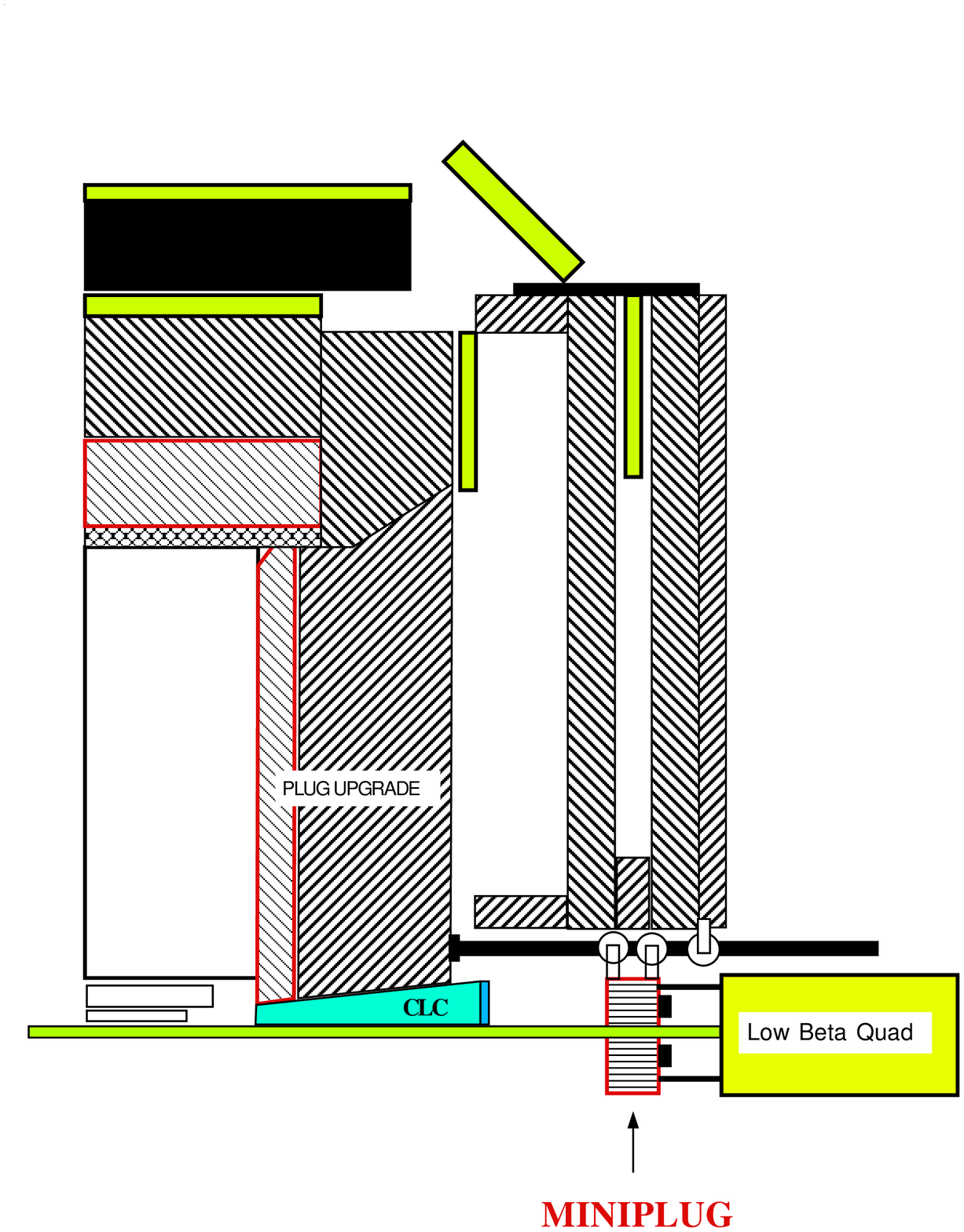,width=1.0\linewidth,angle=0}
%\vspace{-.3cm}
\vglue -0.2cm
{\footnotesize {\bf FIGURE 1.} Schematic drawing of 1/4 view of the 
CDF detector 
showing a MiniPlug hanging from two unistrut beams 
supported on one end by the Plug and on the other 
by the toroid (not to scale). 
This scheme allows for moving the toroids and/or the Plug 
to access the central detector while the MiniPlug remains stationary.}}
\end{minipage}
\setcounter{figure}{1}

\noindent A 28 $X_0$ prototype
of cross section $15\times 15~{\rm cm}^2$
was constructed in 1994 and tested first with  2, 3 and 5 GeV 
electron beams and an 8 GeV $\pi^-$ beam 
at the Brookhaven National Laboratory (BNL) \cite{BNL,NIM}
and then with high 
energy positrons, muons and pions in the 1997 Fermilab test beam. 
The test results
\cite{TUCSON,BRAZIL,nimlast}, summarized in section III,
were found to be in good agreement with simulation
predictions \cite{Elba}.

The final design of the MiniPlugs is described in section IV.
Approximately, the design characteristics of passive and 
active materials were transferred from the prototype to the final design.
In order to reduce the cost of the readout electronics,
a coarser granularity was used,
with a hexagonal cell structure about two times larger than the
4 cm$^2$ of the prototype. However, this was balanced by a higher
number of fibers/cell (6 instead of 4) and a larger fiber diameter (1~mm instead
of 0.83~mm). 
A cosmic ray test
was performed, and its results are presented in section V, while first
collider data are described in section VI.

\section{Design Concept and GEANT Simulation}

\noindent\begin{minipage}[b]{2.5in}{ 
%\vspace*{-.1cm}
\vglue .3in
\hspace*{-2.0cm}
\centering\leavevmode
\epsfxsize=.8\textwidth
\epsfbox [90 240 382 461]{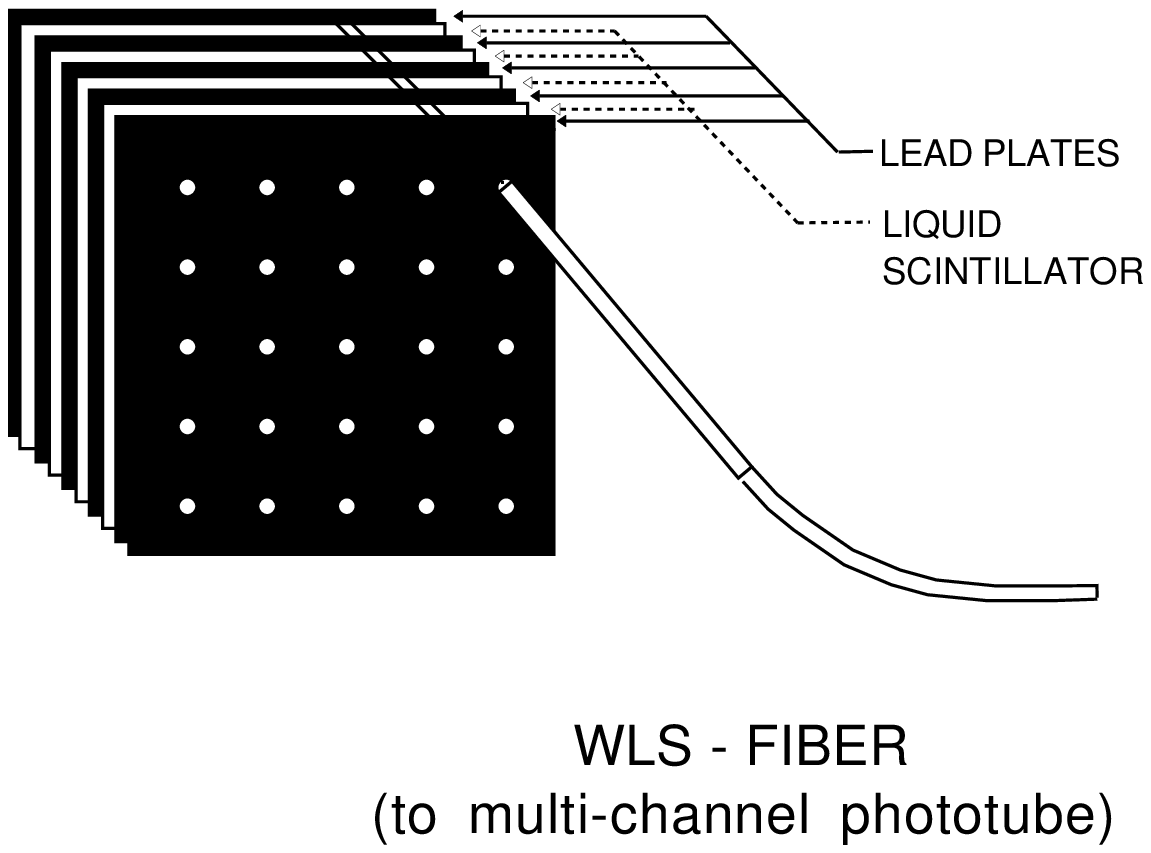}

\vglue -.1in
{\footnotesize {\bf FIGURE 2.} 
Conceptual design of the MiniPlug calorimeter: the fibers
are arranged in a `towerless' geometry, i.e. with no discrete tower
boundaries.}}
\end{minipage}
\setcounter{figure}{2}
\hspace*{.05in}
\begin{minipage}[t]{3.1in}{ 
\vglue -6.3cm
The MiniPlugs consist of alternating layers 
of lead plates and liquid scintillator read out by WLS
fibers. The fibers are inserted into an array of aligned 
holes drilled in the lead 
plates as shown in fig.~2 and  are read out 
by MAPMTs.
Effective `towers' are formed by combining together the desired number
of fibers to be viewed by a single pixel of a MAPMT.

For particles incident at small angles relative to the fiber direction, 
the fiber pulse height centroid provides the position 
of the shower initiating particle.
Position determination 
is obtained not only for electrons and photons}
\vglue .1in
\end{minipage}

\noindent but also for hadrons, 
which upon interaction release on average 1/3 
of their energy in the form of $\gamma '{\rm s}$, mainly from
$\pi^0\rightarrow 2\gamma$ decays, which will interact again after 1 $X_0$.
So a several $X_0$ deep tracking calorimeter can be used, without
a shower maximum detector, to measure the position and energy
of both neutral and charged hadrons. For the MiniPlugs it is not
convenient to be several $\lambda$ long,
to avoid the broadening of the hadron shower if a measurement
of the particle position at the level of few millimeters is required.

We have studied
the response of the MiniPlug calorimeters to single particles and
jets in the
CDF--II configuration by using a simulation based on the
GEANT MonteCarlo program \cite{brun}.
Each MiniPlug module comprised about 56 $X_0$
and 2 $\lambda$. 

\noindent\begin{minipage}[b]{2.7in}{ 
The energy resolution for single particles was analyzed using beams of 
electrons and pions at incident particle energies between 10 and 100 GeV.  
The MiniPlug energy resolution for 
electrons is well described by the formula
$\sigma/E=18\%/\sqrt{E}$, where $E$ is the incident particle energy in GeV.
About $15\%$ of the pions traversed
the calorimeter module without interacting.  
For charged pions 
interacting in the Miniplug the energy resolution was
found to be about $30\%$, independent of the pion energy;
this is dominated by the fluctuation in the ratio $\pi^0/\pi^{\pm}$ in
the first interaction, independent of energy if the hadron shower
is not contained.}
\end{minipage}
%
%\hspace*{.02in}
\begin{minipage}[t]{3.1in}{ 
\vglue -10.cm
%\centering
\vglue .3in
\centering\leavevmode
\hspace*{-0.9cm}
  \epsfxsize=1.0\textwidth
 \epsfbox[ 0 0 567 567]{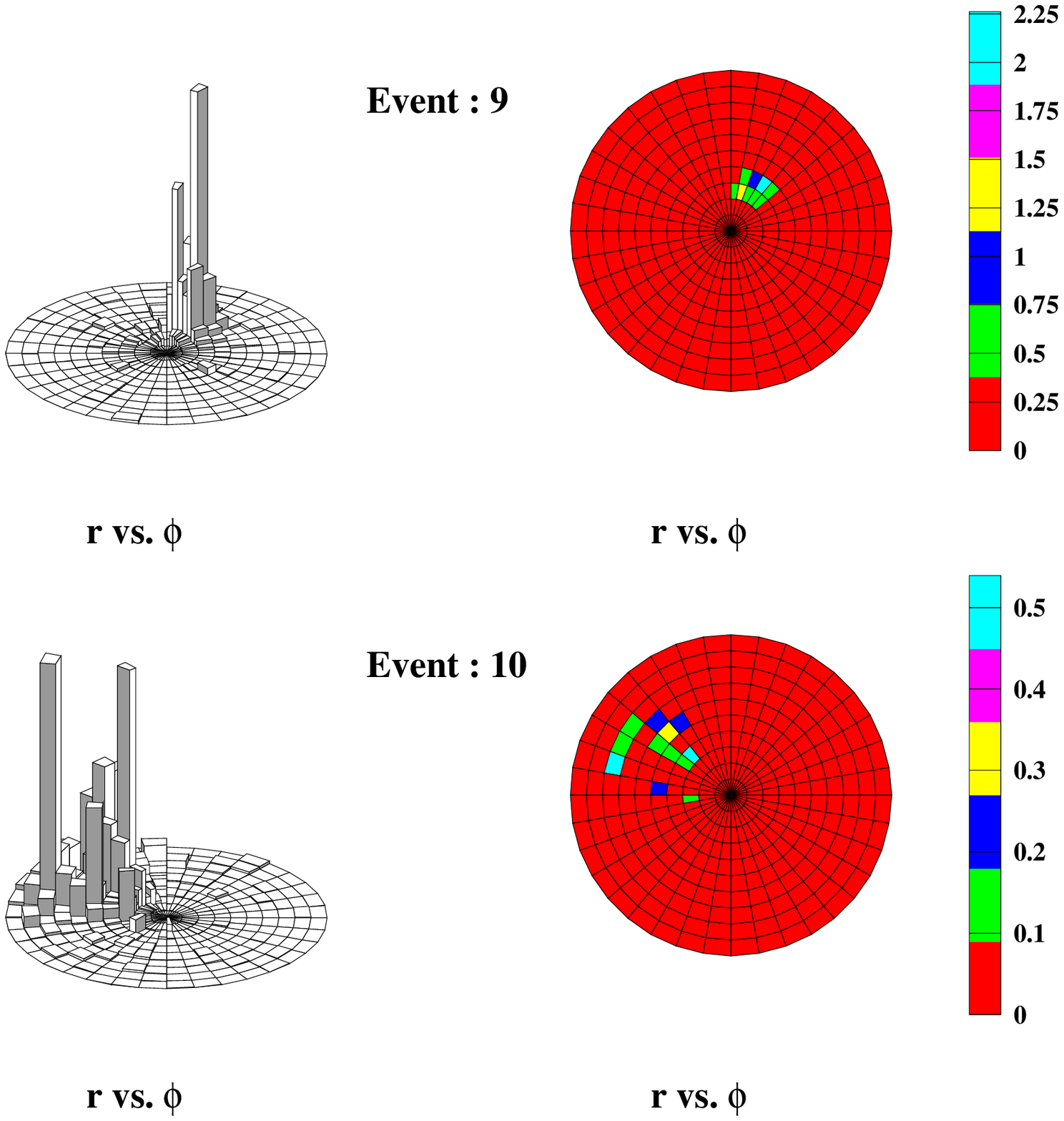}
\vglue -.1in
{\footnotesize {\bf FIGURE 3.} 
MiniPlug response to Monte Carlo generated jets. For  each
event an isometric lego plot (left) and color-coded front view (right) are
shown.}}
\end{minipage}
\setcounter{figure}{3}

Jets were generated using the HERWIG \cite{herwig} MonteCarlo program.
The jet energy resolution was studied by passing individual particles of
a jet through the MiniPlug module and summing the response to each particle
of the jet. 
%This means that the MonteCarlo simulation is taking in account
%that for instance $\pi^{0}'$${\rm s}$ will be absorbed in the
%first half of the detector and charged hadrons will either not interact
%at all, 
%or be sampled for their secondary electromagnetic products.
Fig.~3 shows the

\noindent\begin{minipage}[b]{2.55in}{
 MiniPlug response to typical jets. For each
event an isometric lego plot (left) and  
color-coded front view (right) are
shown. For convenience in representation, the impact point of a particle
was expressed in polar coordinates $r$ and $\phi$ relative to the center
of the MiniPlug front surface.

Figure~4 a) shows the distribution of the incident jet energy of the
generated events.
The distribution of the number of particles in a jet is presented in
fig.~4 b). Figure~4 c) shows the percent fraction
of the initial jet energy registered in the scintillator; from this
distribution we evaluated the jet energy resolution to be $\sigma/E=$ 29.2$\%$.
The $z$-position of
}
\end{minipage}
\begin{minipage}[t]{3.2in}{ 
\hspace*{0.06cm}
\vglue -10.2cm
\centering
\vglue .3in
  \epsfxsize=1.1\textwidth
 \epsfbox[ 0 0 567 567 ]{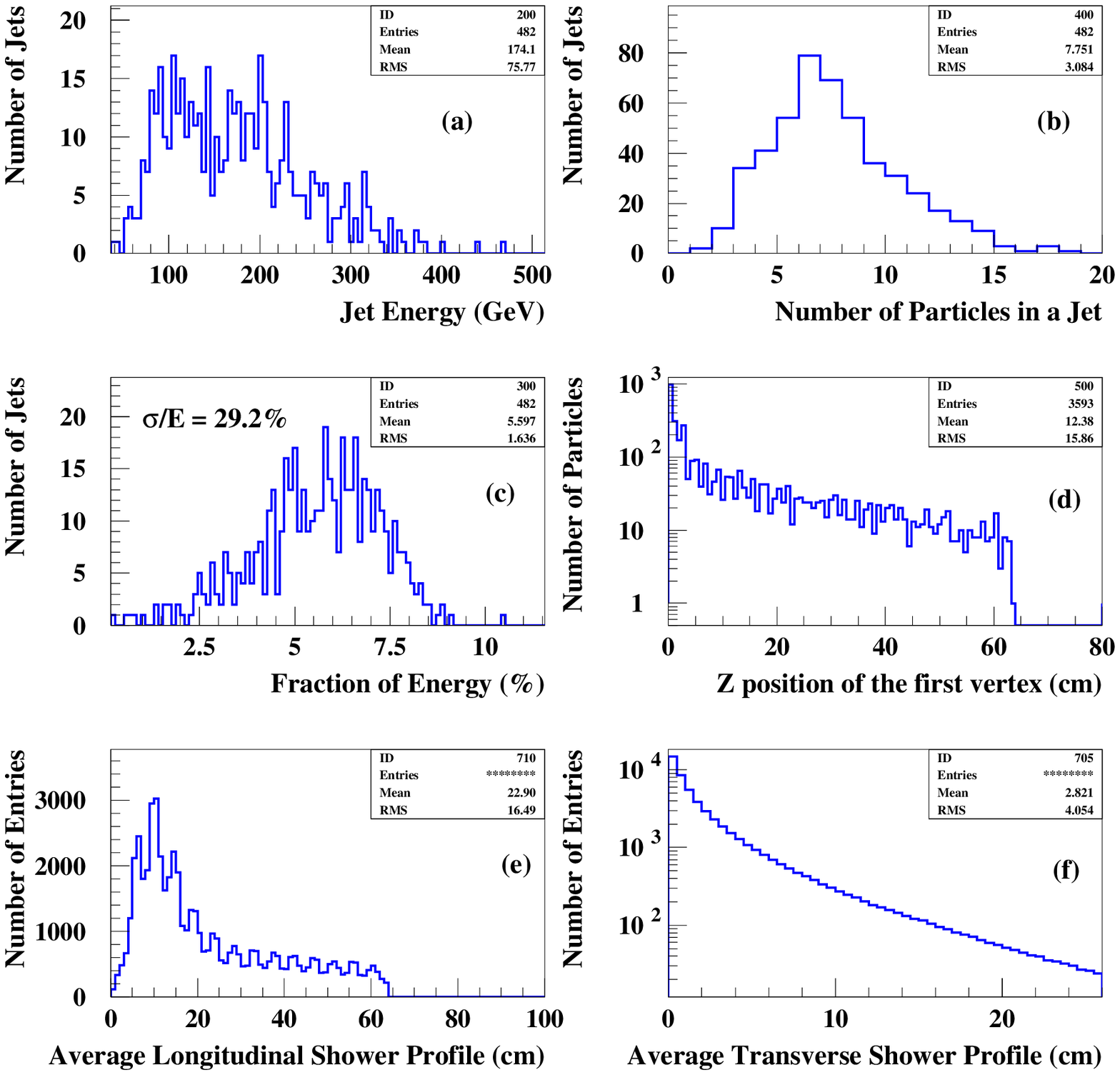}
\vglue -.15in
{\footnotesize {\bf FIGURE 4.}
Distributions for jets incident on a 2$\lambda$ MiniPlug
obtained with a GEANT simulation.}}
\end{minipage}
\setcounter{figure}{4}
\begin{minipage}[b]{3.5in}{ 
\centering
\vglue -.2in
%\hspace*{0.2cm}
\epsfig{file=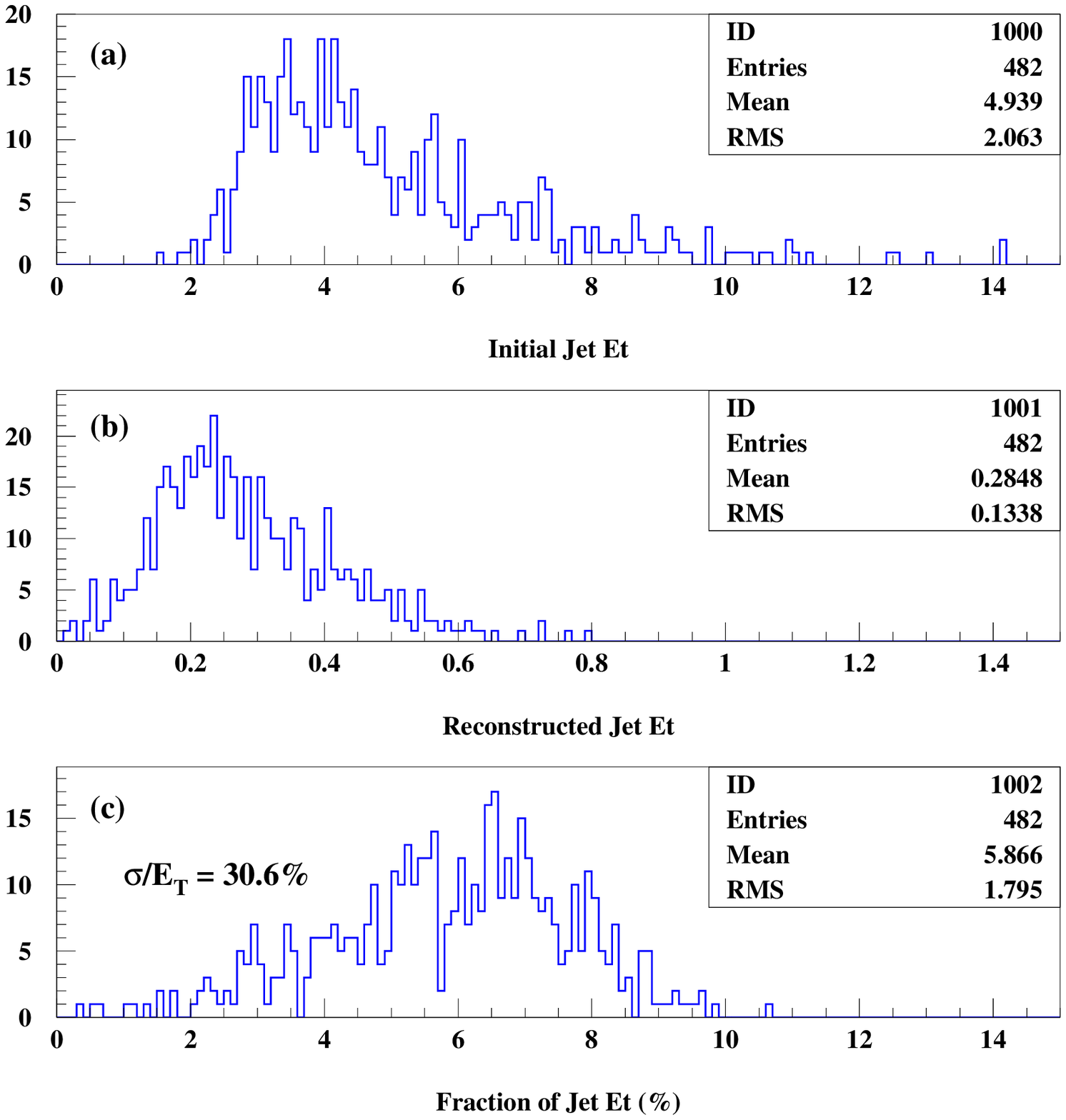,width=1.04\linewidth,angle=0}
\vglue -.1in
{\footnotesize {\bf FIGURE 5.} 
Jet $E_T$ resolution in a 2 $\lambda$ MiniPlug obtained with
a GEANT simulation:
(a) Initial jet E$_T$ distribution, (b) reconstructed
jet E$_T$, (c) fraction
of the initial jet E$_T$ reconstructed in the MiniPlug.}}
\end{minipage}
\setcounter{figure}{5}
\hspace*{.07in}
\begin{minipage}[t]{2.1in}{ 
\vglue -9.9cm
 the first shower vertex,
the average longitudinal and
transverse shower profile are shown
in figures~4 d), 4 e), and 4 f),
respectively.

Results for the jet E$_T$ resolution in the MiniPlug are presented in
fig.~5. The initial jet E$_T$ distribution is shown in 
fig.~5 a),
while the reconstructed jet E$_T$ is plotted in 
fig.~5 b). Figure~5 c) shows the percent fraction
of the initial jet E$_T$ reconstructed in the MiniPlug, which yields
a jet transverse energy resolution of 30.6$\%$ for the
average $E^{jet}_T$ of 4.9 GeV.}
\end{minipage}

\section{The Prototype Performance}
A schematic side view of the MiniPlug prototype is shown in 
fig.~6(left). 
The prototype consisted of 30 parallel lead plates, with dimensions
$15\;{\rm cm}\times 15\;{\rm cm}$ and 4.8 mm thick, spaced 6.4 mm apart.
The plates were laminated with 0.5 mm thick aluminum sheets of 86$\%$
reflectivity glued on the lead with epoxy.
The same lead plate thickness and scintillator gap are used in 
the full scale Miniplugs.
Multiclad WLS Kuraray Y--11(350)M fibers of 0.83 mm diameter
were inserted into an array of aligned holes in the plates through the entire
depth of the detector. The holes have 1 mm diameter and were drilled
with a CNC machine.
The whole plate/fiber assembly was immersed in mineral oil based
Bicron 517L liquid scintillator. The MiniPlug prototype comprised
about 28 $X_0$ or about 1 $\lambda$.
\vglue -.2cm
\begin{minipage}[t]{2.9in}{
\centering
\hspace*{-1.6cm}
\epsfxsize=1.5\textwidth
\epsfbox[31 30 761 582]{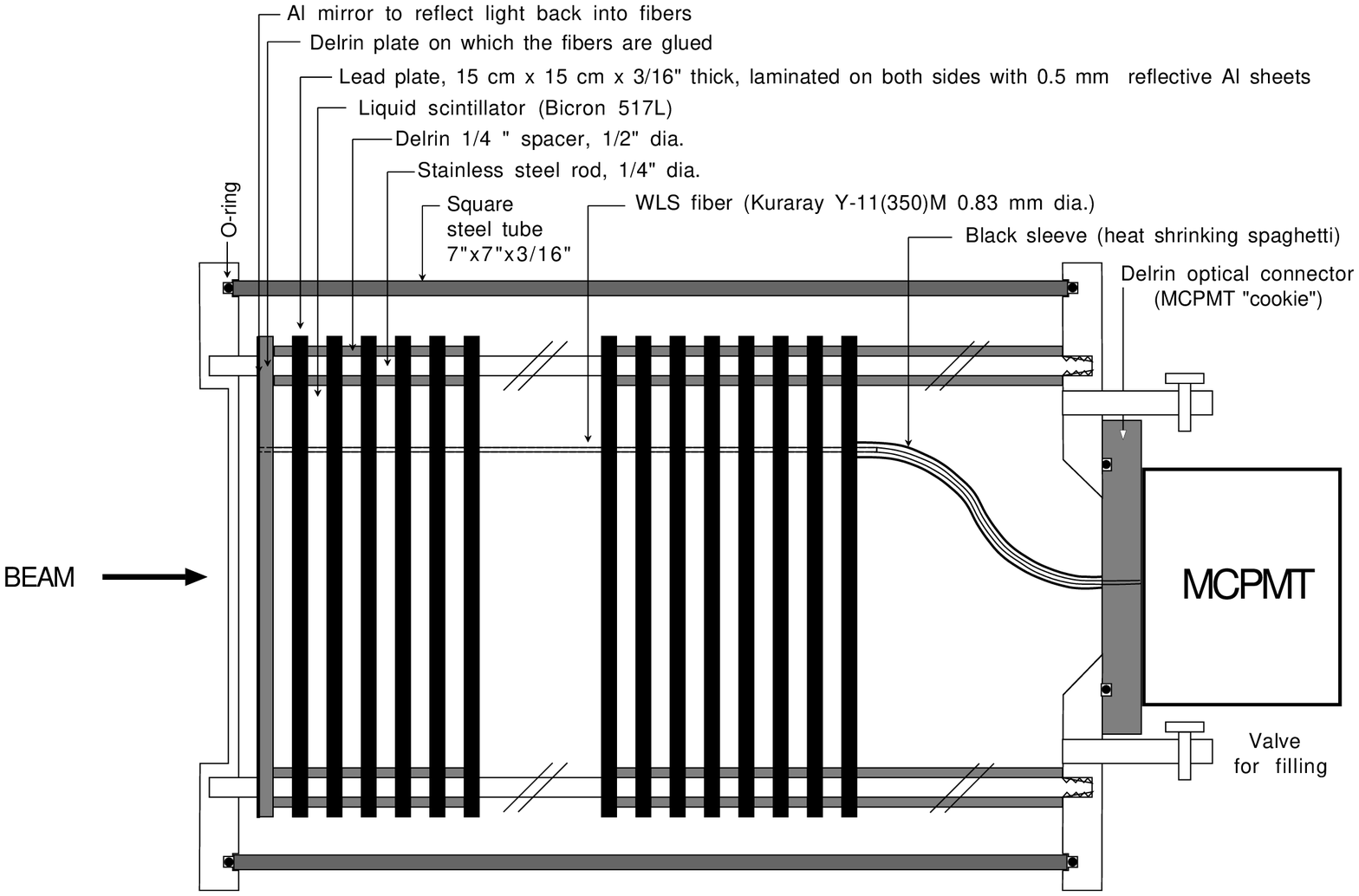}
}
\end{minipage}
\begin{minipage}[t]{2.7in}{
\hspace*{.6in}\vspace*{-7.1cm}
\hspace*{.4in}\epsfxsize=.9\textwidth
\epsfbox[30 30 583 761]{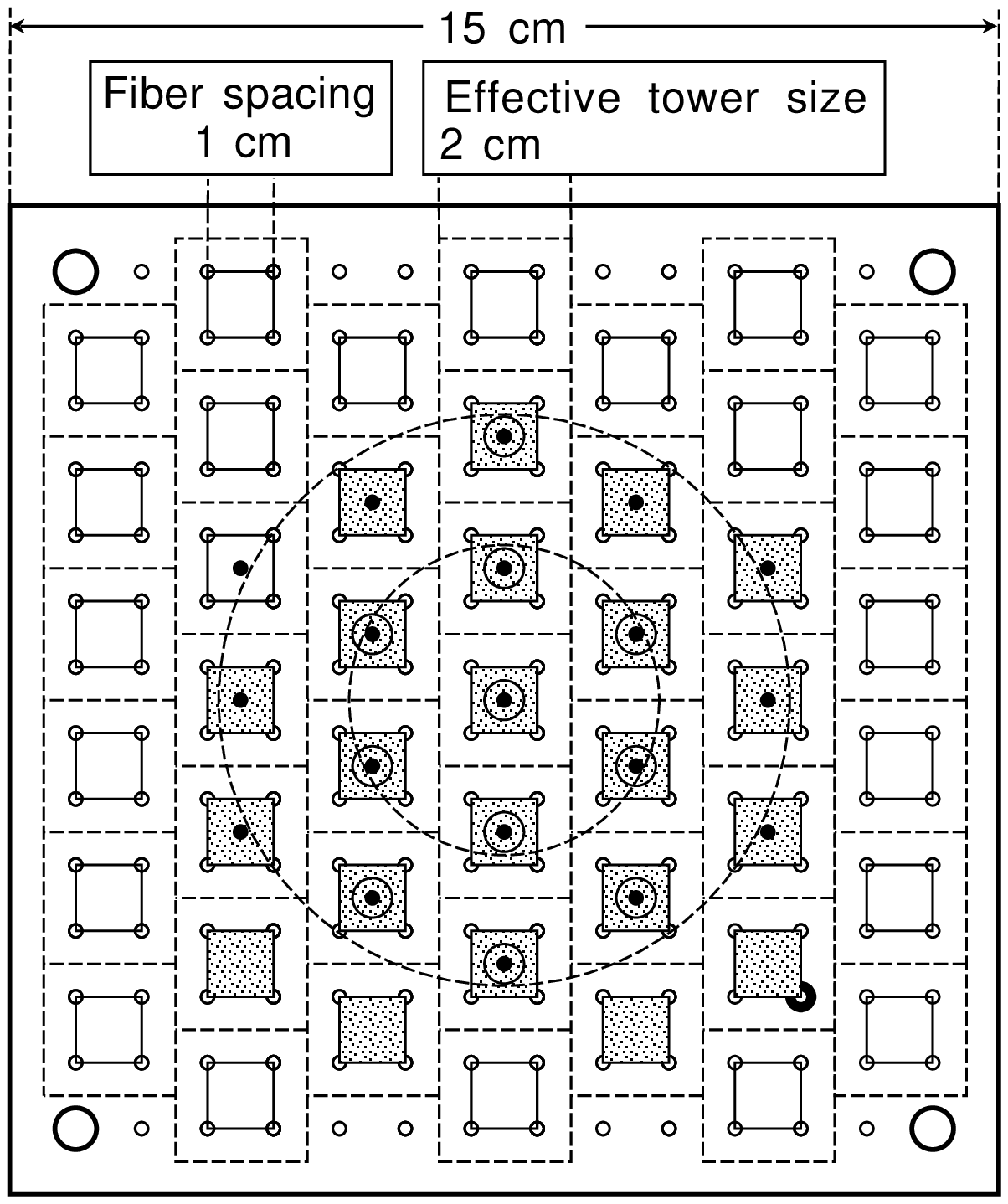}
%\vspace*{.7cm}
\vglue -0.15cm
}
\end{minipage}
\setcounter{figure}{6}
\vglue -2.4cm
\footnotesize{{\bf FIGURE 6.}
Schematic side view (left) and front view (right)
of the MiniPlug prototype.
Each group of four fibers
in a square is read out by a single MAPMT channel. All the squares were
instrumented at the 1997 test beam, as well as the single fibers at the
center of the squares, as indicated (the shaded squares are those instrumented
at the BNL test~[3]).}

\normalsize \vglue 0.1cm
In the prototype, the fibers
were grouped as shown in fig.~6(right), with four fibers
at the corners of a square read out by one MAPMT pixel,
forming towers of effective size $2\times 2$ cm$^2$. 
A reflective aluminum sheet was pressed against the far-ends of
the fibers to reflect the light back into the fibers.
The fibers in the centers of the squares were read individually.
In an optimal design of a calorimeter of $\sim$2$\lambda$
or  $\sim$56$X_0$,
these single fibers 
would penetrate the calorimeter only through a certain distance from the back,
stopping short of reaching the first 24 $X_0$.
In this way they would not be sensitive to electron/photon showers and
therefore would provide a ``{\it hadron tag}''.
In the prototype these fibers were brought all the way to the front
of the calorimeter to measure the expected light yield 
when tagging hadrons with one fiber per
``tower''.

The MiniPlug operates by integrating the signal over several fibers within a
distance determined by the effective
attenuation length of the liquid scintillator, which depends on the number
of reflections.
Bench measurements yielded an effective attenuation length of 
20 mm for a scintillator thickness of 6 mm~\cite{NIM},
which drove the choice of the tower size.

\begin{figure}[htp]
\centering
\vspace*{-.35cm}
\hspace*{-0.4cm}
\epsfxsize=.49\textwidth
\epsfbox[0 0 595 482]{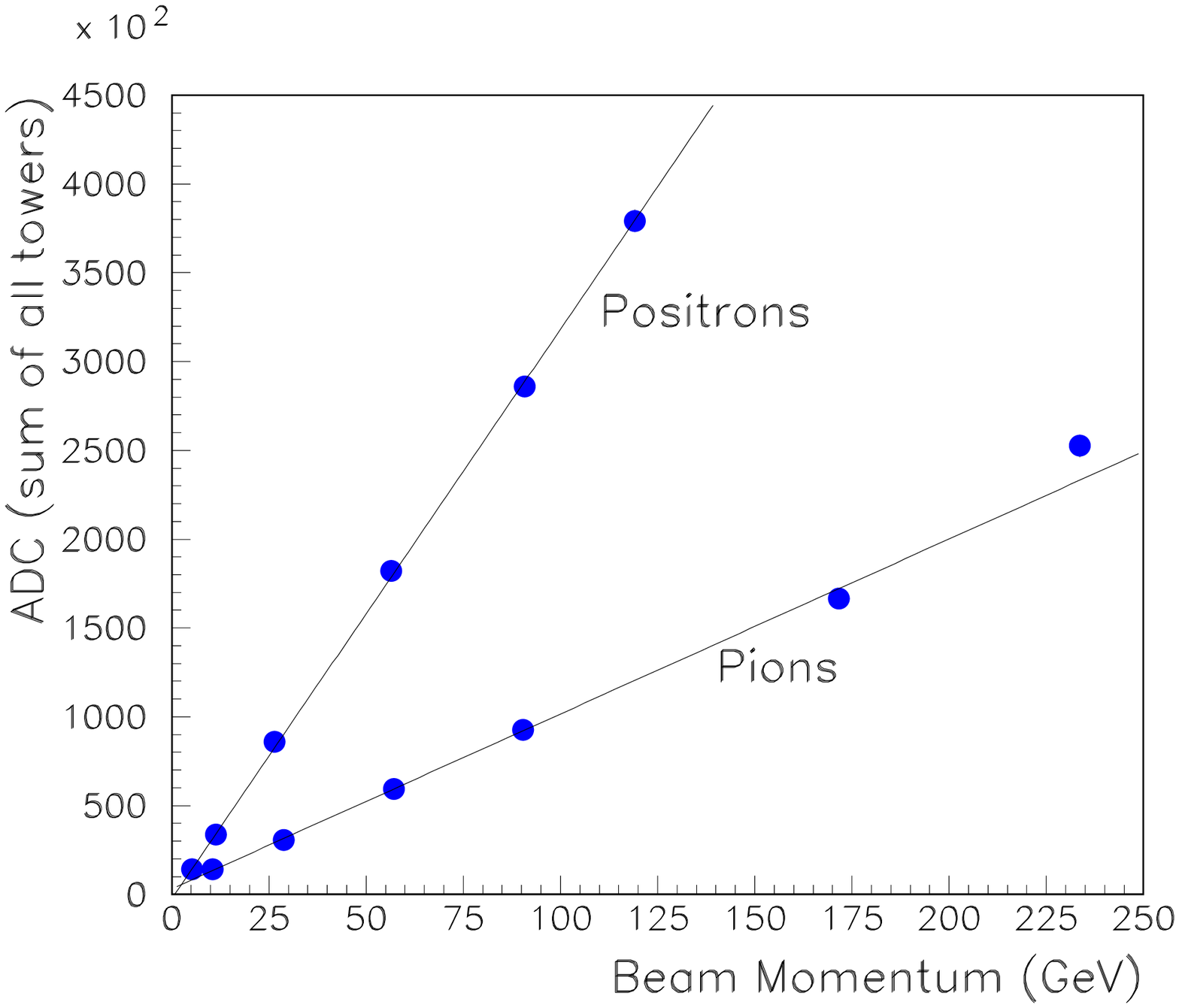}
\epsfxsize=.49\textwidth
\epsfbox[0 0 595 482 ]{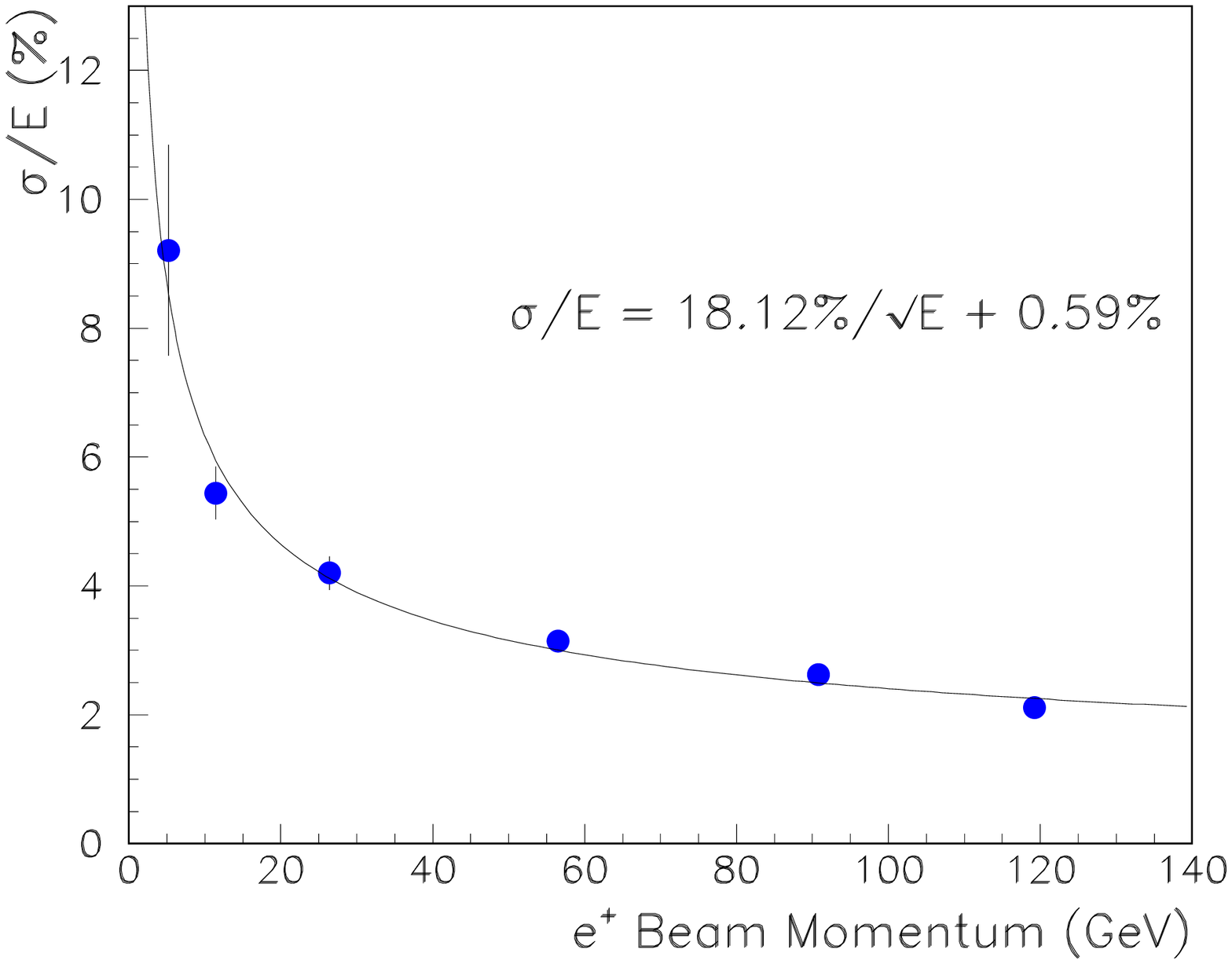}
\vglue 0.02in
\caption{Left: The relation between ADC counts and beam momentum for
positrons in the range 5-120 GeV and pions in the range 10-230 GeV,
obtained in a beam test of the MiniPlug prototype, with
the calorimeter plates perpendicular to the beams.
Right: MiniPlug energy resolution for positrons.}
\vglue -.1in
\label{linearity}
\end{figure}

\begin{figure}[h]
\centering
\vspace*{-.3in}
\hspace*{-0.3cm}
\epsfxsize=.49\textwidth
\epsfbox[0 0 595 482]{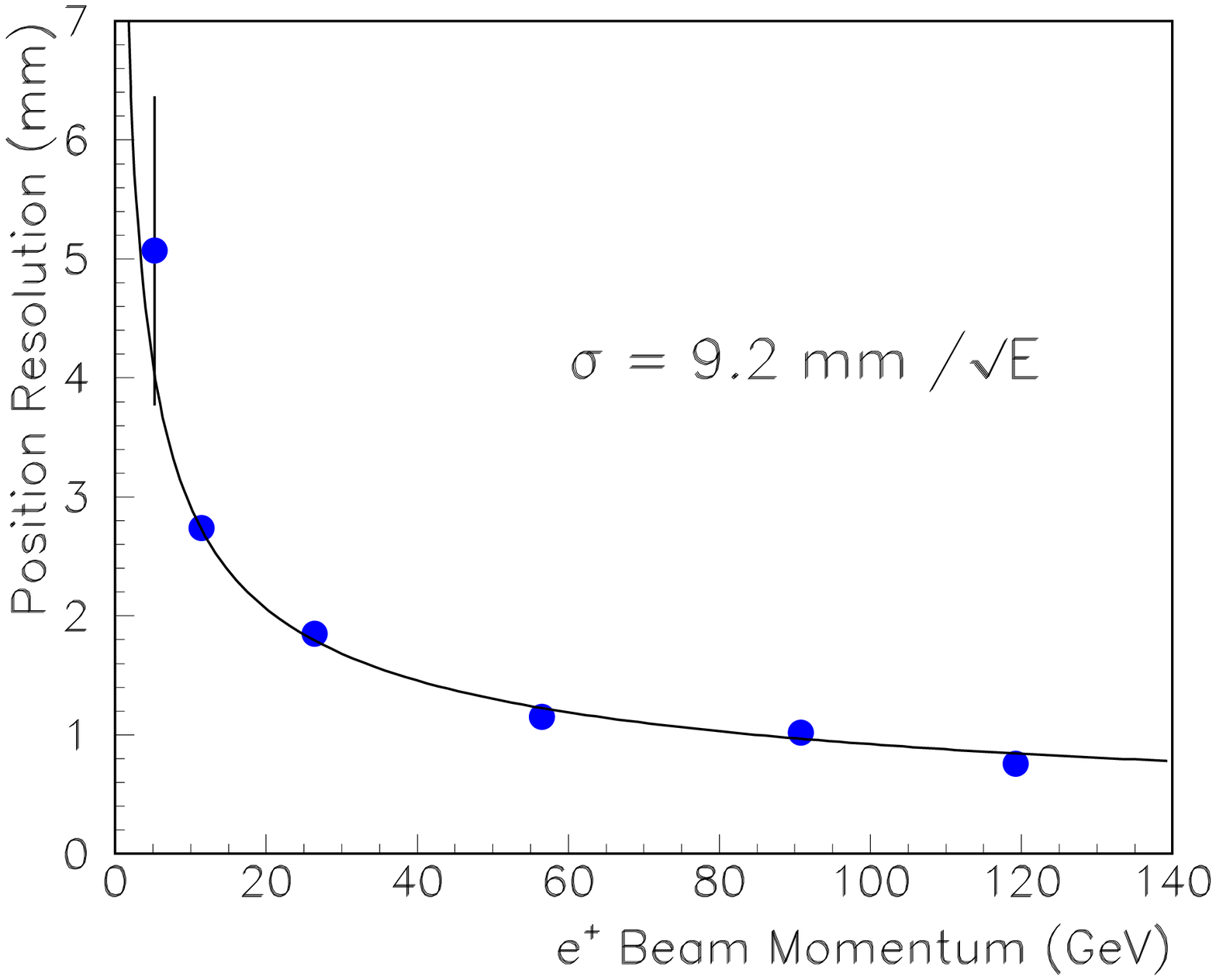}
\epsfxsize=.49\textwidth
\epsfbox[0 0 595 482]{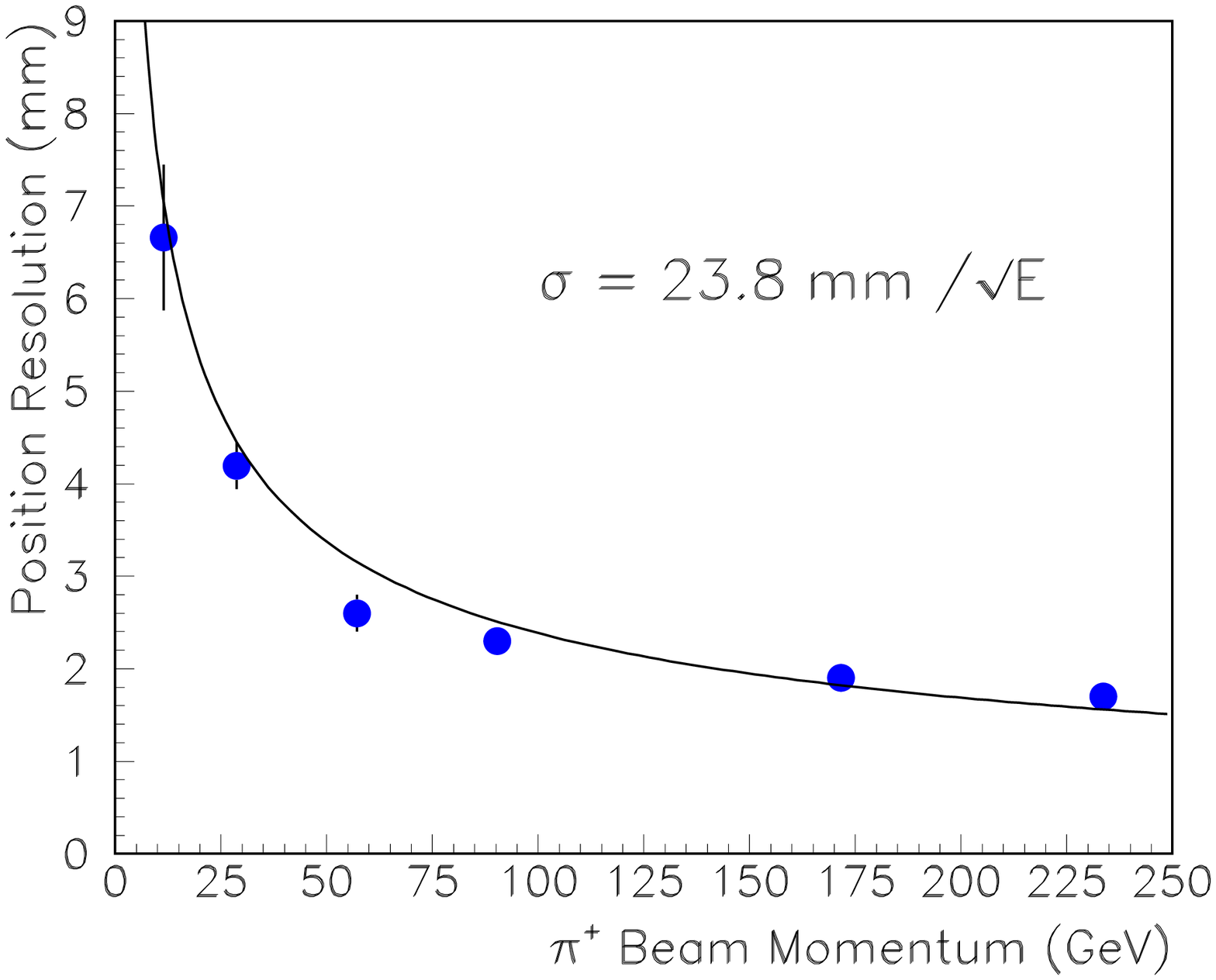}
\vglue 0.03in
\caption{MiniPlug prototype position resolution for positrons in the
range 5-120 GeV (left) and pions in the  range 10-230 GeV (right).}
\vspace*{-.3in}
\label{pos}
\end{figure}

The MiniPlug prototype was exposed to high energy
positron, pion and muon beams at Fermilab in 1997.
Sets of data were taken at angles of 0, 3 and 10 degrees
between the beam and the calorimeter axis.
The response of the prototype to positrons in the
range 5--120 GeV and to charged pions in the range 10--230 GeV showed
very good linearity. For positrons, the deviations from a linear fit were
smaller than 1.5\% (fig.~\ref{linearity}, left) and the average
energy resolution $\sigma$/E~=~18.1$\%$~/~$\sqrt{E}$~+~0.6$\%$
(fig.~\ref{linearity}, right).
For interacting charged pions the energy resolution was about $40\%$, 
independent of the pion energy.  
The lateral position resolution was measured to be 
$9.2\,{\rm mm}/\sqrt{E}$ for positrons and $23.8\,{\rm mm}/\sqrt{E}$ 
for pions (fig.~\ref{pos}).

\section{MiniPlug Final Design}
\begin{figure}[htp]
\centering
\vglue -0.6in
\epsfxsize=.65\textwidth
\epsfbox[ 31 31 583 761]{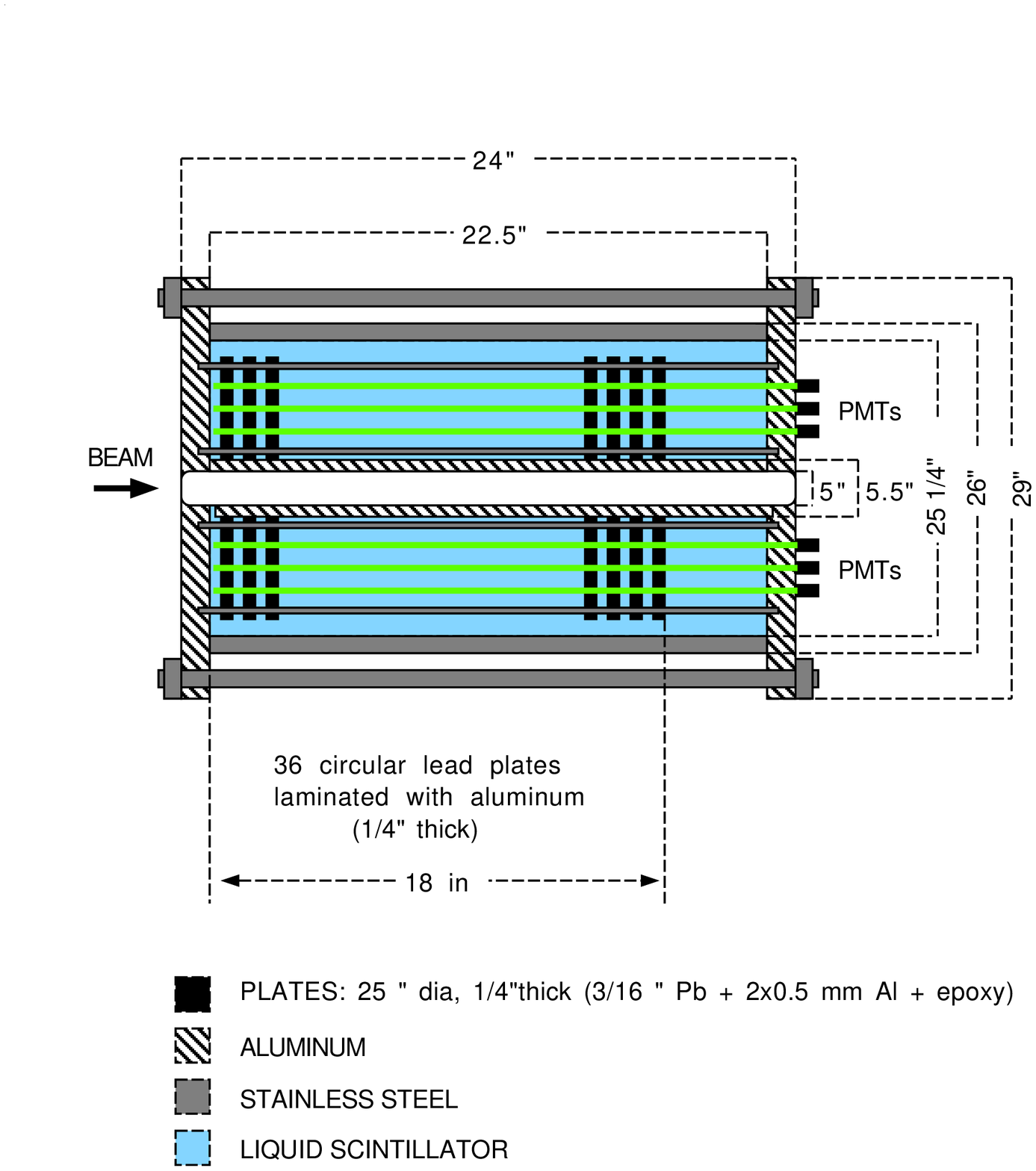}
\vglue -1.in
\caption{Schematic side view of a Miniplug (not to scale).}
\label{MP_CDF2_side}
\vglue -.1in
\end{figure}

A schematic drawing of a side view of a MiniPlug 
is presented in fig.~\ref{MP_CDF2_side}. 
The active calorimeter consists of 36 lead-based plates  
separated by 1/4$''$ spacers and immersed in 
liquid scintillator. 
Table~\ref{tab:params} summarizes various parameters of
the MiniPlug calorimeters.

\begin{table}[b]
\vglue -.1in
\renewcommand{\arraystretch}{0.95}
\caption{Parameters of the MiniPlug calorimeter.}
\begin{tabular}{l|c} \hline\hline
    Sensitive length              & 571.5 mm  \\
    Inner sensitive radius & 139.7 mm   \\
    Outer sensitive radius & 641.4 mm    \\
    Weight &  840 Kg \\
    Depth (including Al flanges) & 609.6 mm\\
    Thickness                  & 32 $X_0$, 1.3 $\lambda$   \\
    Number of plates & 36   \\ 
    Plate thickness    & 4.8mm (Pb+6\% Sb) + 1mm Al\\
    Scintillator gap thickness  & 6.4 mm \\
    Scintillator        & Bicron 517L \\
    Wavelength shifter  & Kuraray Y11 1mm dia. \\
    Phototube & Hamamatsu R-5900 M16 \\
    Number of WLS fibers/pixel & 6 \\
    Number of phototubes & 18 \\ \hline\hline
\end{tabular}
\label{tab:params}
\end{table}

The calorimeter plates were manufactured by 
Alchemy Castings Inc. \cite{welco}.
They are 25$''$ (63.5 cm) in diameter and are
made of 3/16$''$ thick (Pb$+6\%$ Sb) alloy laminated with 0.5 mm Al
of 80\% reflectivity.
Each plate has 1512 holes for the fibers, drilled with a 
$\#$54 (or 1.4 mm) drill and champhered. 
In addition, 12 holes of 1/4$''$ and 6 holes of 1/2$''$ nominal diameter 
are used for `hanging' the plate in its final position in the calorimeter from
stainless steel rods supported by the MiniPlug end plates,
as seen in  fig.~\ref{MP_CDF2_side}.
A MiniPlug plate is 0.86 $X_0$ and 0.03 $\lambda$ thick, and weighs 21 Kg. 
The active length of a MiniPlug, including the liquid scintillator, 
is 32 $X_0$ and 1.3 $\lambda$, and 
the total weight, including the liquid scintillator, 
is approximately 840 Kg (1850 lbs).
Because of the mentioned space limitations, the final design
has no hadron tagging fibers.

\begin{figure}[htp]
%\begin{center}
\vglue -0.1in
\centering\leavevmode
  \epsfxsize=0.58\textwidth
 \epsfbox[ 30 33 582 761 ]{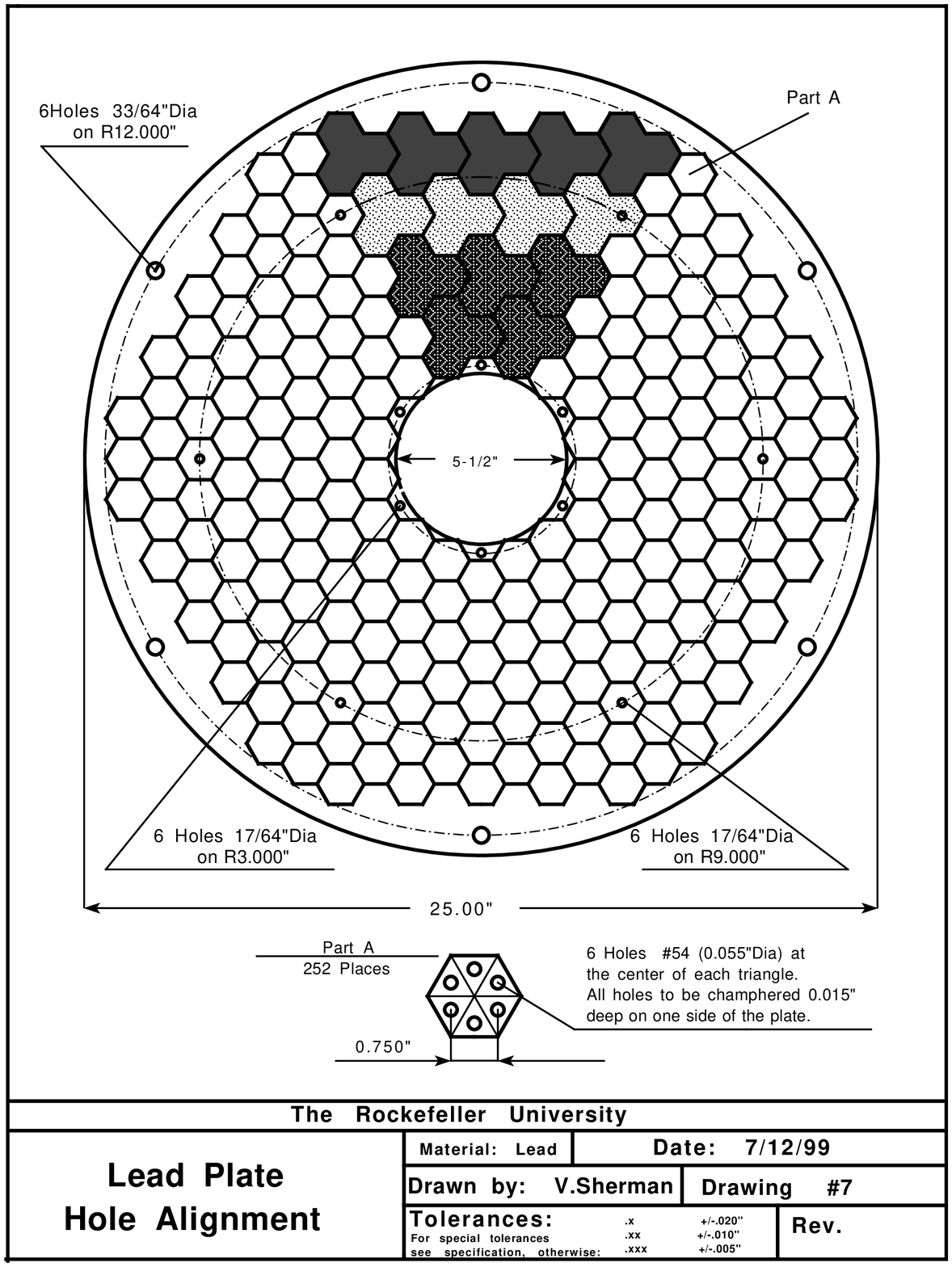}
\vglue .05in
\caption{MiniPlug lead plate.
The tower design is based on a hexagon geometry. Each hexagon has six holes, with
a WLS fiber inserted in each hole. 
The six fibers of a hexagon are grouped together and are
viewed by one MAPMT channel. 
There are 252 hexagons in each MiniPlug viewed by 18 16-channel MAPMTs.
The MAPMT outputs are added in groups of 3 to form 84 
calorimeter `towers'.}
\label{fig:mp_vadim1}
\vglue -.1in
%\end{center}
\end{figure}

Figure~\ref{fig:mp_vadim1} shows the layout of the fiber holes in a Miniplug plate. 
The design is based on a hexagon geometry. Each hexagon has six holes, as shown.
A WLS fiber is inserted in each hole. 
The six fibers of a hexagon are grouped together and are
viewed by one MAPMT channel. A seventh fiber, which is clear and carries
the light from a calibration LED, is also read out by each MAPMT pixel.
There are 252 hexagons in each MiniPlug viewed by 18 16-channel MAPMTs.
The MAPMT outputs are added in groups of 3 to form 84
calorimeter `towers'. The single hexagonal cells could 
eventually be read out individually.
Presently we read out 84 towers per MiniPlug and 
the summed MAPMT outputs (18 channels per MiniPlug), amounting to a total
of 204 channels for the two MiniPlugs.
The towers viewed by the three MAPMTs of a sextant are shown in 
fig.~\ref{fig:mp_vadim1} with different shades.

We use Bicron 517L liquid scintillator \cite{BC517L} 
and Kuraray Y11 (350) multiclad 1.0 mm dia.
WLS fibers read out by Hamamatsu R5900-M16 MAPMTs \cite{R5900-M16}.

The Bicron 517L liquid scintillator, also used in the prototype,
is a mineral oil with Pseudocumene as the active ingredient.
Because of its high chemical compatibility with polystyrene based fibers,
low cost and radiation hardness, the BC-517L scintillator had 
been considered and studied extensively for applications at the 
Superconducting Super Collider (SSC).

In the MiniPlug prototype we used one 0.83 mm dia. fiber per cm$^2$, and 
grouped four fibers together to form a `tower' of effective 
area 4 cm$^2$.  In the final MiniPlug design,
there is one fiber per 1.6 cm$^2$, but the 1 mm dia. of the fibers 
compensates for the reduction in light expected due to the larger area covered 
by each fiber. Moreover, the fibers are now aluminized at the `far end', while
in the prototype they were pressed against a reflective plate.
%The aluminization of the fiber ends was done at Fermilab. 
%Then, at Rockefeller, 
%the aluminized ends were covered with a bead of Epoxy, which serves a 
%double purpose: protect the aluminization and prevent the fibers 
%from accidentally being pushed through the last lead plate during 
%the Miniplug assembly process.

\vglue .15in
\begin{minipage}[b]{2.7in}{ 
\centering\leavevmode
\epsfxsize=.75\textwidth
 \epsfbox[ 14 14 90 115 ]{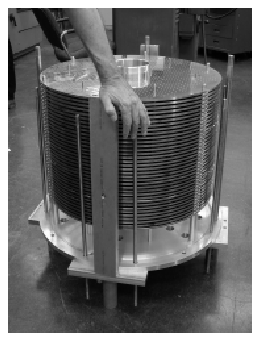}

{\footnotesize {\bf FIGURE 11.} 
A dry stack of the plates of a  MiniPlug.}}
\end{minipage}
\setcounter{figure}{11}
\hspace*{.1in}
\begin{minipage}[t]{2.7in}{
\vglue -7.8cm
\centering\leavevmode
\epsfxsize=0.75\textwidth
\epsfbox[  14 14 90 115 ]{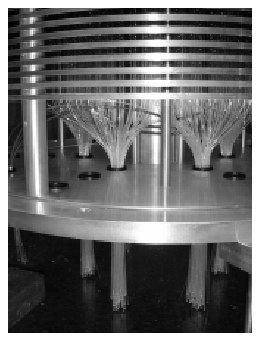}

{\footnotesize {\bf FIGURE 12.} 
A view of the fiber routing.}}
\end{minipage}
\setcounter{figure}{12}

\vglue 0.05in
The Hamamatsu R5900--M16, a 16 channel tube, was selected  
because of 
its relatively low cost per channel and acceptable
channel to channel gain uniformity. 

The expected radiation dose in the MiniPlugs in Run II has been estimated to be 
$\sim$ 30 KGy (3 Mrad) per year. Such a dose is not expected 
to affect the response of the liquid scintillator.
SSC studies on WLS fibers
showed that the permanent damage in light transmittance 
of 1.0 mm dia., 1 m long 
fibers is about 10$\%$ after an irradiation of 64 KGy \cite{SSC}.
Radiation hardness studies performed for the Large Hadron Collider (LHC)
 showed that
Kuraray Y11 fibers were the least damaged when compared to other commercially
available fibers \cite{LHC}. 
The standard borosilicate glass of MAPMT windows
is  sensitive to radiation and, when tested in connection with 
SSC applications, showed a
60$\%$ relative transmittance at 500~nm for a 30 KGy dose.
For this reason, the R5900 MAPMTs for the Miniplugs were ordered with 
a quartz window, which substantially improves the MAPMT radiation hardness.

\vglue 0.05in
\noindent\begin{minipage}[b]{2.8in}{
The two MiniPlug calorimeters have been assembled at Rockefeller.
Fig.~11 shows a dry stack of the plates of a
MiniPlug, and a view of the fiber routing is shown in
fig.~12.
After the fibers have been sealed into the optical connectors
on the back plate, they have been cut
with a diamong string-saw, and the 
connector surface hand polished.
Fig.~13 shows one MiniPlug at Fermilab before
installation. On the back plate are mounted 18 MAPMTs, plus two LED
assemblies for calibration.} 
%A seventh fiber, which is clear and carries
%the light from one of these calibration LED's, is also read out by
%each MAPMT pixel.
\end{minipage}
\hspace*{.1in}
\begin{minipage}[t]{2.7in}{
\vglue -6.1cm
\centering\leavevmode
  \epsfxsize=1.0\textwidth
 \epsfbox[14 14 157 121 ]{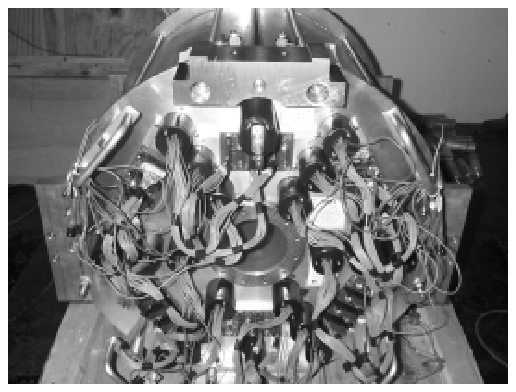}

\vglue .1in
{\footnotesize {\bf FIGURE 13.} 
A MiniPlug at Fermilab before installation.}}
\end{minipage}
\setcounter{figure}{13}

\section{Cosmic Ray Test}
We tested a $60^{\circ}$ section of one MiniPlug using
cosmic ray muons.
In the first step we 
determined the channel response 
uniformity (and the 
operating voltage) of the tubes to be used on the MiniPlugs.
Then the single photoelectron response  
was measured using a
${}^{60}$Co gamma source. Finally,
a cosmic ray trigger was made of a 2-fold coincidence of two 
scintillation counters mounted above and below the MiniPlug module.
The front-end readout system employed for these tests was a CAMAC based 
LeCroy 2249W~\cite{LC2249}
11-bit 12-channel charge integrating ADC, with
a sensitivity of 0.25 pC/count,
controlled by a 
National Instrument `Labview'~\cite{Labview} data acquisition software.

\subsection{Photomultiplier Test}
\label{sec:pmt_test}

The light carried by the WLS fibers is read out by
16-channel Hamamatsu H6568-10/R5900 M16 tubes~\cite{R5900-M16},
whose cross section has a size of about one square inch and
whose specifications are listed in Table~\ref{tab:pmt_specs}.
In addition to the 16 anode channels, these MAPMTs have a
channel which is the sum of the last dynodes of all 16 channels.
The output of this channel will be referred to as the `sum-output'.

\begin{table}
\renewcommand{\arraystretch}{0.97}
\caption{Characteristics of the Hamamatsu H6568/R5900 M16 photomultiplier
    tubes specified for the MiniPlug calorimeter.}
\begin{tabular}{l|c} \hline\hline
    Parameter       & Description/Value    \\ \hline
    Window material              & quartz  \\
    Quantum efficiency at 500 nm & $\ge$13~\%   \\
    Anode dark current           & 1~nA    \\
    Pulse linearity per channel ($\pm2$\% deviation) & 0.5~mA \\
    Cross-talk                   & 1~\%    \\
    Uniformity between each anode& 1:2.5   \\ \hline
    Photocathode material         & bialkali\\
    Spectral response            & $300 \sim 650$~nm \\
    Number of stages             & 12               \\
    Anode                        & array of $4\times 4$ \\
    Pixel size                   & 4~mm $\times$ 4~mm \\
    Maximum high voltage         & 1000~V \\
    Anode pulse rise time        & $\le1$~nsec \\
    \hline\hline
\end{tabular}
\label{tab:pmt_specs}
\end{table}

\begin{minipage}[b]{2.8in}{
\vglue -.15in
\centering\leavevmode
\epsfxsize=.58\textwidth
\epsfbox[ 103 6 515 782]{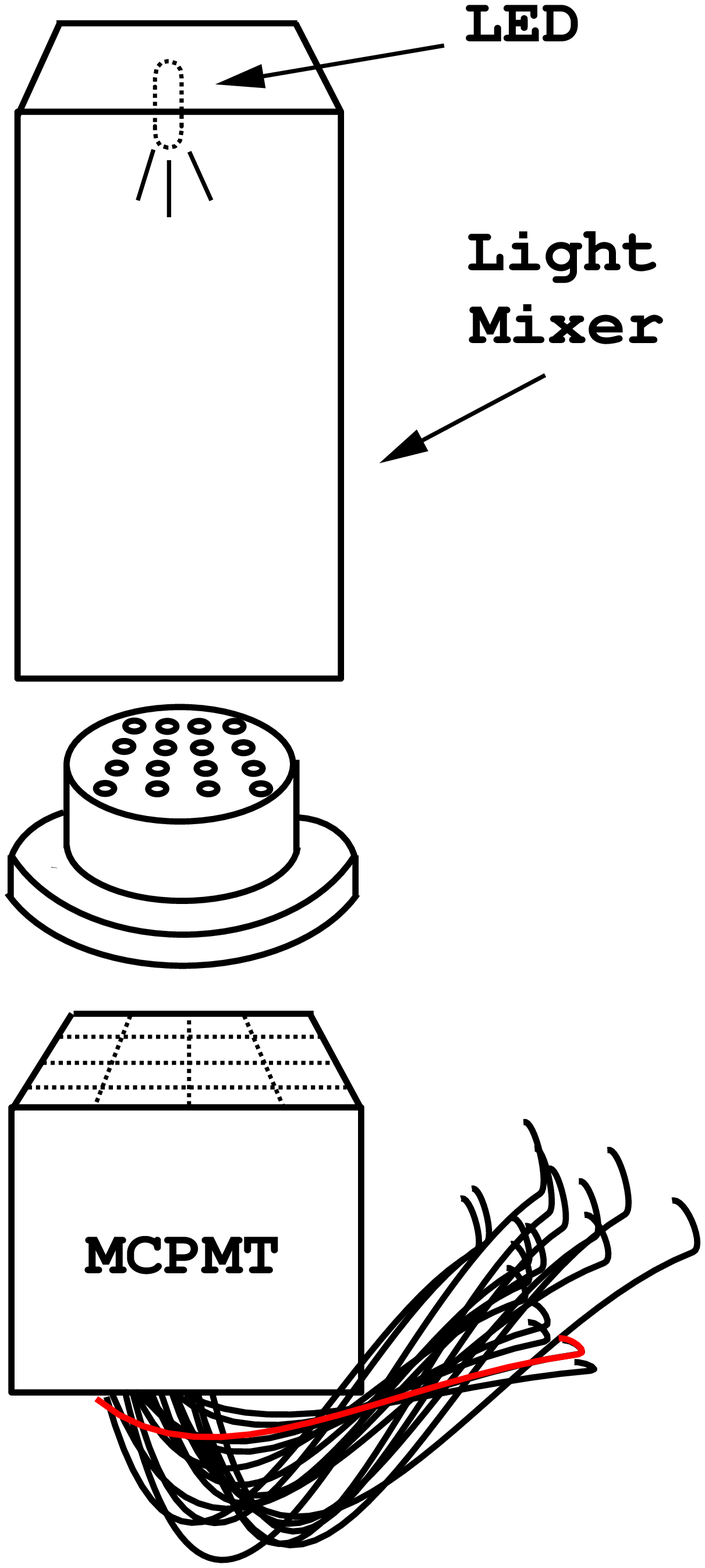}

{\footnotesize {\bf FIGURE 14.} 
The setup of the MAPMT test system uses a green LED 
embedded in a light mixer to distribute the light uniformly into the 
16 channels of the MAPMT.}}
\end{minipage}
\setcounter{figure}{14}
%
%\hspace*{.1cm}
\begin{minipage}[b]{2.8in}{
\vglue -4.8cm
The setup we used for measuring the relative response of the MAPMT channels
is shown in fig.~14. 
A green LED is placed in a hole drilled at one end of a rectangular 
lucite bar, which acts as a `light mixer' 
distributing the LED light uniformly among 
a set of clear fibers mounted in holes drilled in a delrin `cookie' in 
a pattern matching that of the MAPMT photocathode pixels.
There are 7 fibers in each cookie hole, just as in the final MiniPlug design.
The uniformity of the LED light across the $4\times4$ cookie channels
has been measured with three MAPMTs,
and the measurement has been repeated after rotating each tube 
by $90^{\circ}$, $180^{\circ}$ and $270^{\circ}$.
 Since at each rotation 
a given MAPMT pixel is illuminated by a different cookie channel,
the differences among ADC counts  
}
\end{minipage}

\noindent obtained at the four settings 
reflect the differences in illumination of the corresponding 
cookie channels.
These measurements show a standard deviation of $\pm3$~\%.
Finally, the operating voltage of the tubes has been determined by
equalizing the sum-outputs.
The desired gain for the tubes operating in the MiniPlugs is $10^5$ which,
according to the manufacture's specifications, is achieved at 
a MAPMT voltage of approximately -700V. 
The LED light was adjusted to bring the sum-output of one PMT, whose 
voltage was set to yield a gain of $10^5$, to 450 ADC counts.
Then, for each one of the other tubes, the high voltage was varied in steps 
of $\pm 10$ V in the range from -600 to -800V to 
find the value at which its sum-output was closest to 450 ADC counts.
Variations around the average value -700V were about $\pm5$~\%.

The response of the 16 channels of each MAPMT to the illumination of the 
LED in the setup of fig.~14 was measured and 
compared with the specifications supplied by the manufacturer.   
In the MiniPlugs, the MAPMT channels has been hard-wired into groups of 3,
forming 5 `grouped channel' triplets
and one singlet per MAPMT (the latter, channel $\#16$,
 will not be read out 
in the MiniPlugs).
Therefore, a special connector was made to group the MAPMT channels 
in the test setup, and the response of the 5 grouped channels was also 
measured.
\begin{minipage}[b]{2.4in}{
\noindent The test results for one MAPMT are shown in fig.~15:
(a) relative response of the 
16 channels, defined as the ratio of the 
pulse height of a channel to the maximum pulse height of all 16 channels;
(b) measured relative channel response
compared to values obtained from the manufacturer; 
(c) relative response of grouped channels; and 
(d) the difference between
the relative response of a grouped channel and 
the expectation from the measurement
of the 16 individual channels.

\noindent The `uniformity' index of the 45 tested MAPMTs, 
defined as the ratio of the lowest 
to the highest channel response, showed an average value of 2.1$\pm 0.3$.

Since the sum-output will be used as a trigger, 
it is also important to study the correlation
 between the
}
\end{minipage}
\hspace*{.02in}
\begin{minipage}[t]{3.25in}{
\vglue -11.1cm
\centering\leavevmode
  \epsfxsize=0.98\textwidth
 \epsfbox[20 150 575 670]{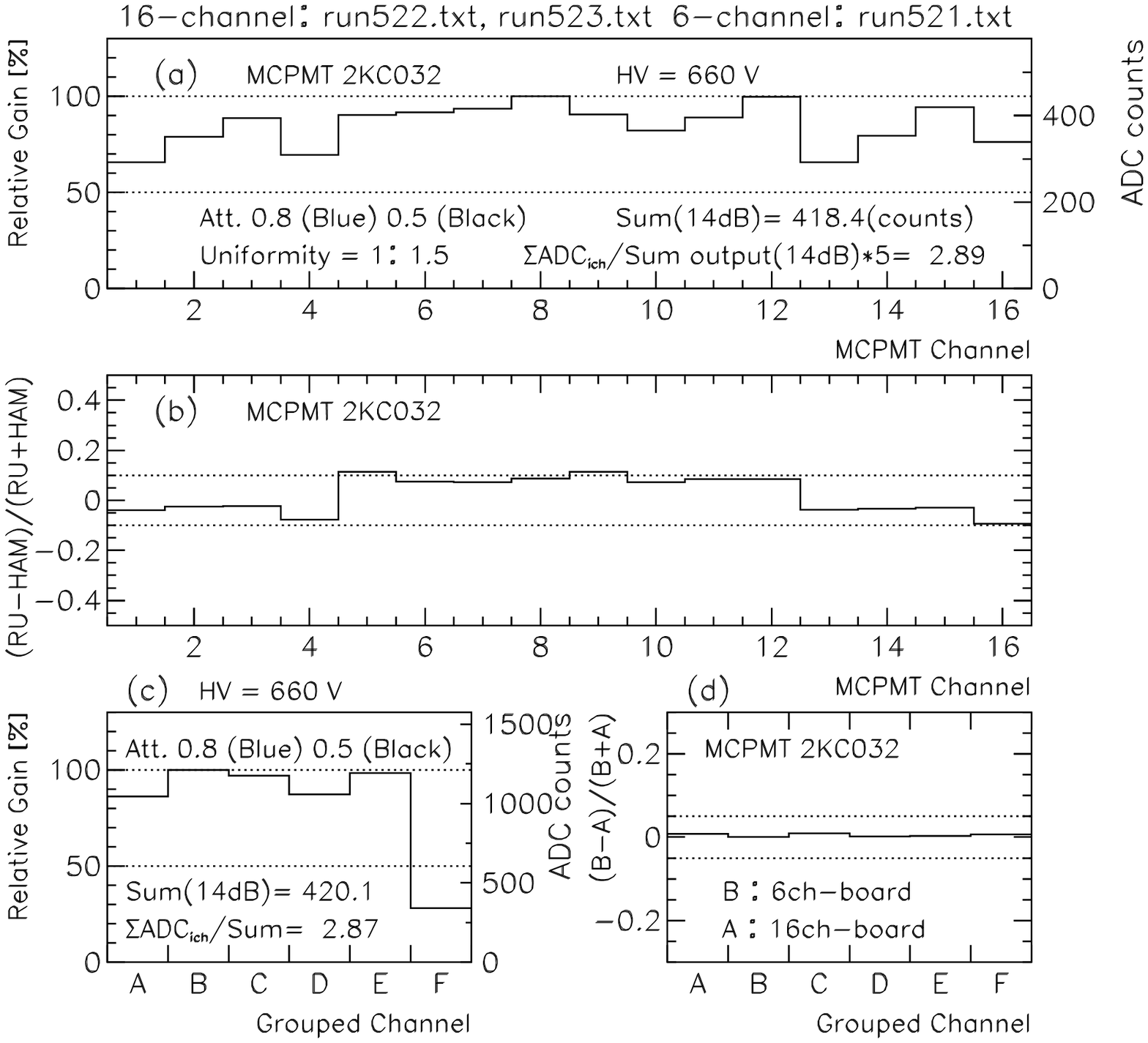}

{\footnotesize {\bf FIGURE 15.} 
(a) Relative response of the 16 channels of one MAPMT; 
(b) difference between
  measured relative response and values provided
  by the manufacturer; (c) relative response of grouped channels; (d) 
  difference between relative response  of grouped channels and 
  the expectation from the measured response of the individual channels.}}
\end{minipage}
\setcounter{figure}{15}

\noindent sum of outputs from the 16 
individual channels
and the sum-output.  
We found that
the ratio of the sum of outputs from the 16 individual channels to the
sum-output is constant in the range of about $10\sim1000$ pC. 

\subsection{Single Photoelectron Measurement}

A measurement of the signal from single photoelectrons 
was carried out using a
${}^{60}$Co gamma source, which
was placed just outside the MiniPlug
vessel. The light emitted by the liquid scintillator 
%(Bicron 517L~\cite{BC517L}) in the MiniPlug 
when irradiated by the source is rather weak and usually results in single 
photoelectron emission by the MAPMT photocathode.
We triggered events using a randomly-generated gate
of width 1 $\mu$s. 
Most triggers do not contain any signal within the gate, but a small fraction 
of events have a single photoelectron signal.

\begin{figure} 
\vglue -.2in
\centering\leavevmode
\epsfxsize=.8\textwidth
\epsfbox[ 28 392 537 650]{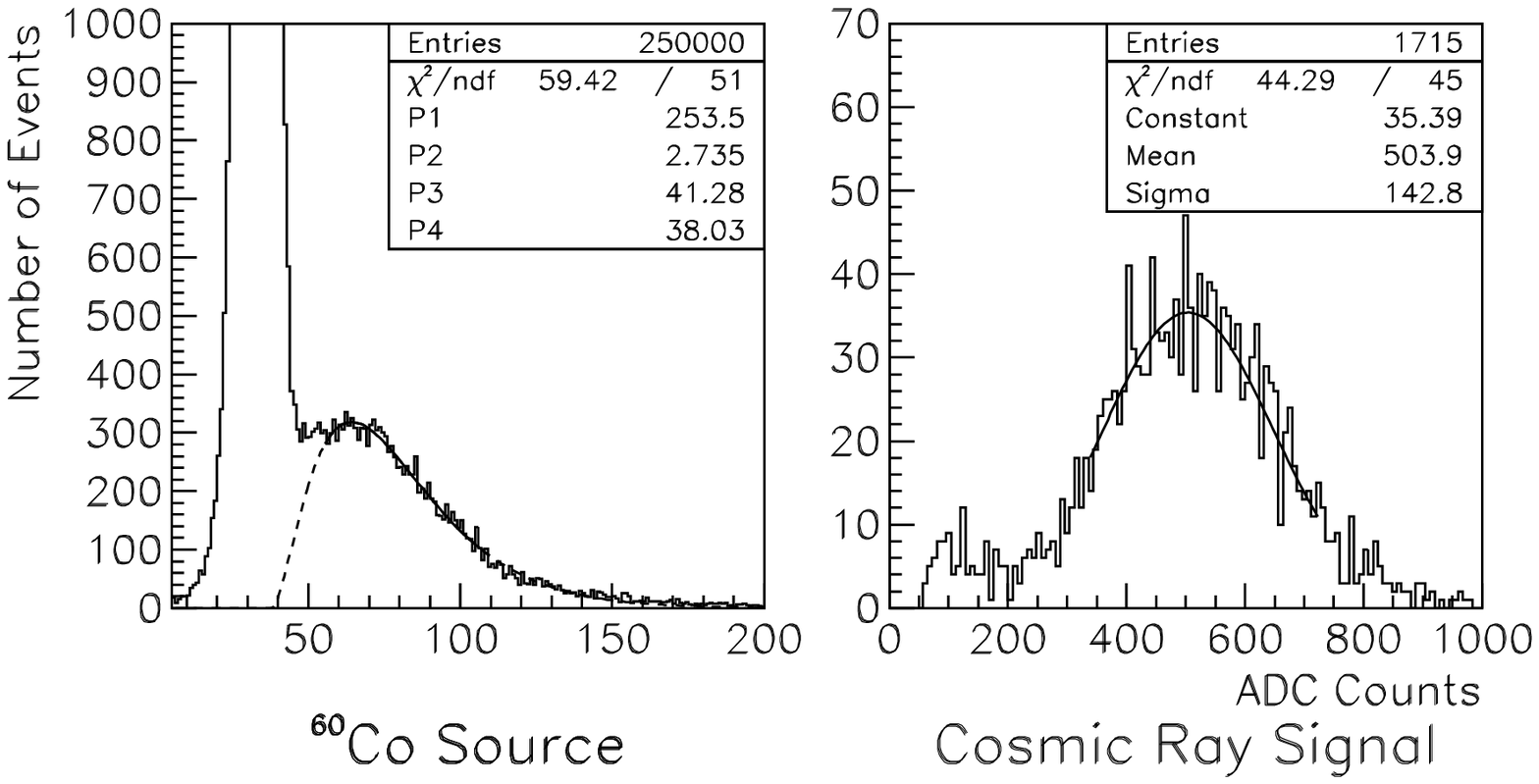}
\vspace*{0.04cm}
\caption{Left: the ${}^{60}$Co source signal for a single tower.
The data are fitted with a {\it Polya distribution}, 
$P(m)=[m(mG/G_0)^{m-1} \cdot e^{-mG/G_0}]/\Gamma(m)$.
The parameter $p3=G_0$ corresponds to the pulse height for single 
photoelectrons.
Right: the cosmic ray spectrum with the isolation cut
fitted with a Gaussian function.}
\vglue -.15cm
\label{fig:test_summary}
\end{figure}

Three MAPMTs to read out one $60^{\circ}$ section of one MiniPlug 
were operated at -960V in this cosmic ray test, corresponding
to a gain of $\sim$5.0$\times10^6$ as specified by the manufacturer, 
At this gain, 
the expected single photoelectron response of the LeCroy 2249W ADC unit 
is approximately 3.2 counts. A
LRS 234 linear amplifier~\cite{LRS234} was used to amplify the signal 
by a factor of 10.

Figure~\ref{fig:test_summary}(left) shows the ${}^{60}$Co source signal
for one tower, with a clear single photoelectron peak.
The distribution has been fitted with a 
{\it Polya distribution}, i.e.
$$
P(m)=\frac{m(mG/G_0)^{m-1}}{\Gamma(m)}\cdot e^{-mG/G_0}
$$
which is the appropriate one for describing single photoelectron 
pulse height distributions. In the fit, $p1$ is the normalization constant, 
$p2$ the parameter $m$, $p3$ the parameter 
$G_0$, and $p4$ the pedestal mean value.
We obtain 41.28/10=4.1 ADC counts for 
the single photoelectron response of this tower.
Similar results were obtained from the other tested towers.

\subsection{Cosmic Ray Test Results}

The cosmic ray trigger was made of a 2-fold coincidence of two 
scintillation counters mounted above and below the MiniPlug module.
The scintillation counters were read out by RCA 8575 PMTs. 

A better separation of the cosmic ray muon peak 
from the pedestal is obtained by plotting the tower response after imposing
an isolation cut which requires that the adjacent towers 
have a signal smaller than their own cosmic ray muon signal. 
Figure~\ref{fig:test_summary}(right) shows the signal distribution
after the isolation cut.
Even with the isolation cut, the cosmic ray muon path 
length differences within a given tower still contribute to the 
width of the muon peak. 
So, although a cosmic ray muon spectrum is expected to follow a Landau 
distribution, the measured distribution is a sum
of several different Landau distributions and is better described by a Gaussian
function to obtain the ADC counts for the cosmic ray muon peak, 
as shown in fig.~\ref{fig:test_summary}(right).
The mean value of the cosmic ray signal distribution for the tested
towers has been measured to be about 400 photoelectrons.
Dividing these values by the corresponding single photoelectron ADC counts 
we obtain on average more than 100 photoelectrons/MIP for 
a single tower.

\section{First Collider Data}

The MiniPlugs were installed in the
the CDF apparatus few months ago, and were operated
stably and reliably since then. However,
they are not fully instrumented yet.
At the time of writing, only the sum-output
(the 17th channel) of each tube is read out.
This means a coarser granularity for the present,
equivalent to 18 towers per MiniPlug.
Figure~17(left) shows,
as a $\phi-\eta$ lego-plot, the average
ADC counts from a Collider run.
The inner ring of towers,
at higher pseudorapidity, has larger signals.
That is in agreement with results obtained with the MBR Minimum-Bias event 
generator~\cite{mbr}, followed by a parameterization of the detector
response,
as shown in fig.~17(right).

\vglue .1in
\begin{minipage}[b]{2.8in}{
\hspace*{-0.1cm}
\centering\leavevmode
  \epsfxsize=.8\textwidth
 \epsfbox[ -1 10 612 541]{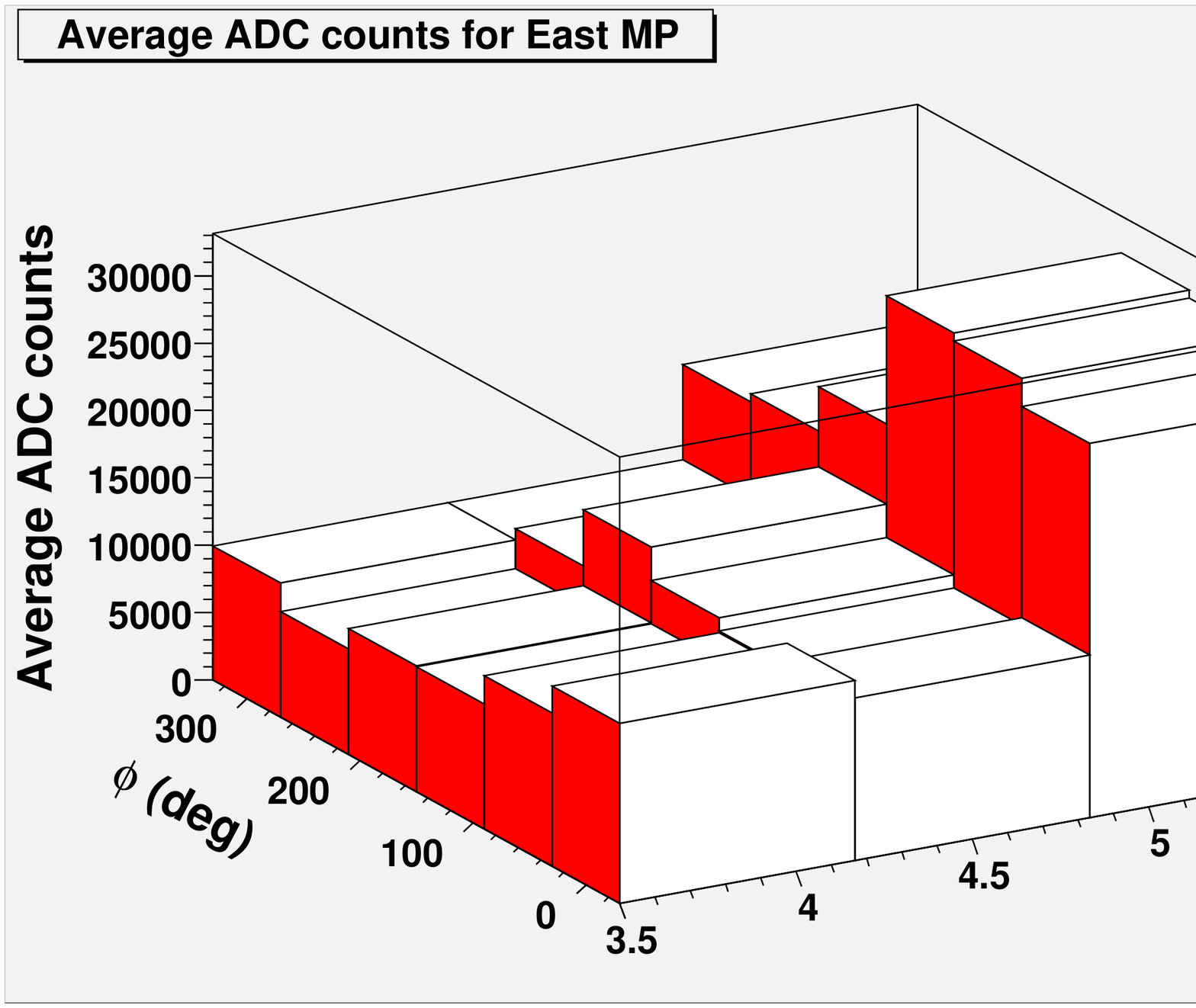}}
\end{minipage}
\begin{minipage}[t]{2.8in}{
\centering
\vglue -5.6cm
\epsfig{file=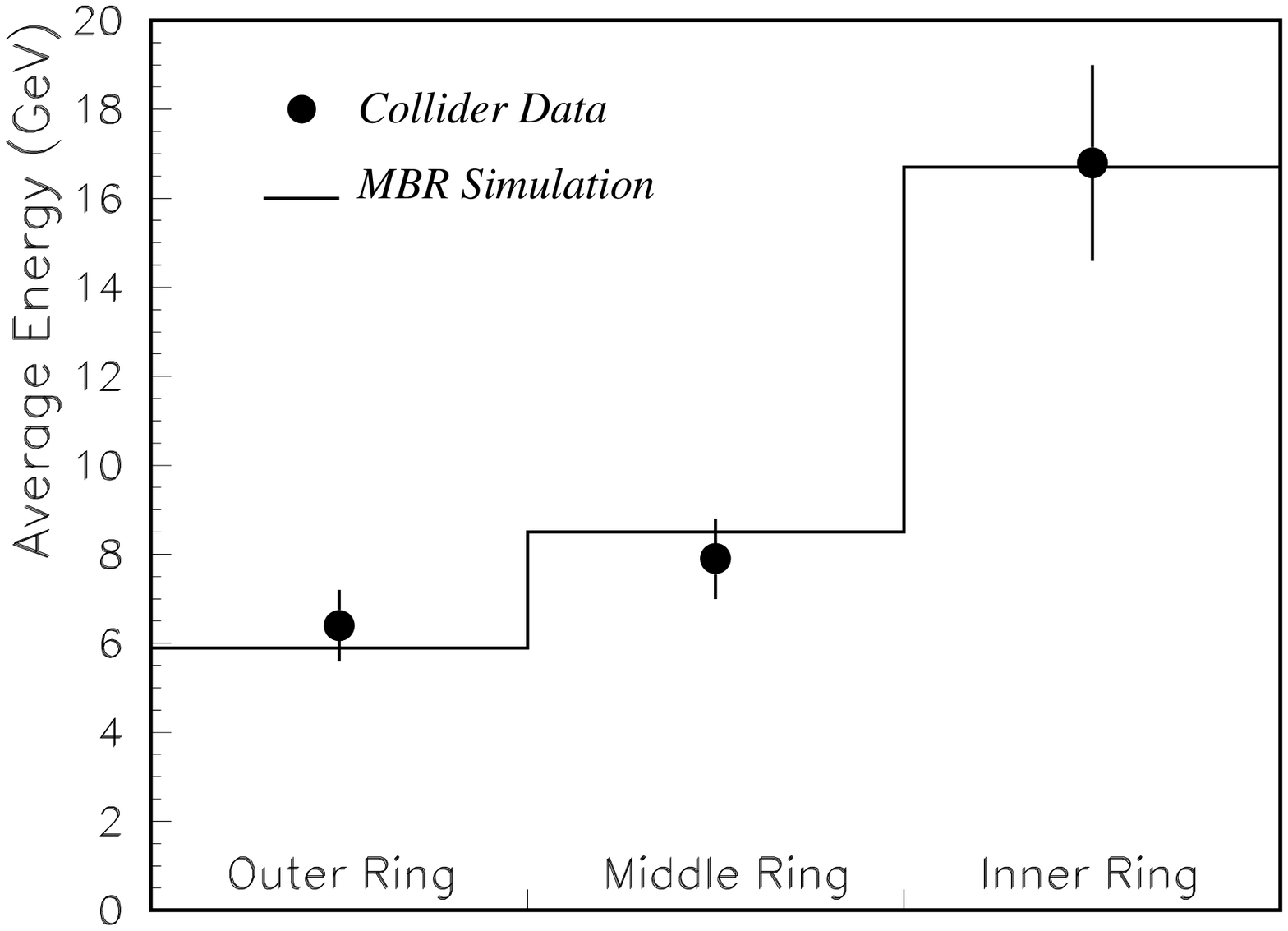,width=1.08\linewidth,angle=0}}
\end{minipage}

\vglue -.2in
\noindent\footnotesize{{\bf FIGURE 17.}
Left: the average ADC counts from a Collider run, for
the 18 towers of a MiniPlug currently read out.
Right: Collider data are compared to an MBR simulation for the
three different pseudorapidity rings.}
\vglue -.2in

\normalsize
\section{Conclusions}
Two MiniPlug calorimeters,
designed to measure the energy and lateral position of particles 
in the forward region ($3.6<|\eta|<5.2$) of the
CDF detector, have been recently installed as part of the Run II CDF upgrade
at the Tevatron $\bar pp$ collider.
They consist of lead/liquid scintillator layers read out by WLS fibers
arranged in a pixel-type towerless geometry. 
In this paper we described the final design of the MiniPlugs and presented
results from a cosmic ray test, in which a light yield of approximately 
100 pe/MIP was obtained, exceeding our design requirements.
First Collider data show a detector response in agreement
with expectations.

\end{document}